# Effect of 50 MeV $Li^{3+}$ irradiation on structural and electrical properties of Mn doped ZnO

**S. K. Neogi**[1], **S. Chattopadhyay**[2], **Aritra Banerjee**[1], **S. Bandyopadhyay**[1], **A. Sarkar** [3] **and Ravi Kumar**[4]

[1] Department of Physics, University of Calcutta, 92 Acharya Prafulla Chandra Road, Kolkata 700009, West Bengal, India
[2]Calcutta Institute of Engineering and Management , 24/1A Chandi Ghosh Road, Kolkata 700040, West Bengal, India
[3]Department of Physics, Bangabasi Morning College, 19 Rajkumar Chakraborty Sarani, Kolkata 700009, West Bengal, India
[4] Department of Material Science & Engineering, NIT, Hamirpur-177005, Himachal Pradesh, India

E-mail: arbphy@caluniv.ac.in (Aritra Banerjee)

**Abstract**

The present work aims to study the effect of ion irradiation on structural and electrical properties and their correlation with the defects in $Zn_{1-x}Mn_xO$ type system. $Zn_{1-x}Mn_xO$ (x = 0.02, 0.04) samples have been synthesized by solid-state reaction method and have been irradiated with 50 MeV $Li^{3+}$ ions. The concomitant changes have been probed by x-ray diffraction (XRD), temperature dependent electrical resistivity and positron annihilation lifetime (PAL) spectroscopy. XRD result shows single phase wurtzite structure for $Zn_{0.98}Mn_{0.02}O$, whereas for $Zn_{0.96}Mn_{0.04}O$ sample an impurity phase has been found apart from the usual peaks of ZnO. Ion irradiation dissolves this impurity peak. Grain size of the samples found to be uniform. For $Zn_{0.98}Mn_{0.02}O$, the observed sharp decrease in room temperature resistivity ($\rho_{RT}$) with irradiation is consistent with the lowering of FWHM of the XRD peaks. However for $Zn_{0.96}Mn_{0.04}O$, $\rho_{RT}$ decreases for initial fluence but increases for further increase of fluence. All the irradiated $Zn_{0.98}Mn_{0.02}O$ samples show metal-semiconductor transition in temperature dependent resistivity measurement at low temperature. But all the irradiated $Zn_{0.96}Mn_{0.04}O$ samples show semiconducting nature in the whole range of temperature. Results of room temperature resistivity, XRD and PAL measurements are consistent with each other.

## 1. Introduction

Dilute magnetic semiconductors (DMS) have attracted much of research interest because of their potential application in the field of spintronics [1,2]. Ferromagnetism (FM) has been achieved both in II-VI and III-V semiconductors by addition of 3d transition metal (TM) elements [3,4]. The euphoria started following the prediction of room temperature ferromagnetism (RTFM) in Mn doped ZnO by Dietl et al. [5]. Thereafter, very often researchers reported intrinsic FM in TM doped ZnO [6-10]. However, the results are quite contradictory regarding the origin of FM in the host semiconductor [1,10-13]. It is also conjectured that defects plays a crucial role in controlling the magnetic properties of such systems [7,8,14-16]. Apart from magnetic properties, structural, optical and electrical properties of TM doped ZnO is also attracting lot of research interest [15,17-19].

Energetic ion beam irradiation is an efficient tool for introducing defect states in solid materials. Consequently, it is an important technique for controlled modifications of structural, optical and magnetic properties of semiconductors [20,21]. There are reports of irradiation studies on ZnO, both with light [22-24] and heavy [25,26] ions. But there are only limited reports of ion irradiation effects on TM doped ZnO. Ravi Kumar et al. reported RTFM and a metal-semiconductor transition in 200 MeV $Ag^{15+}$ ion irradiated ZnO thin films implanted with Fe and observed that oxygen vacancies and/or Zn interstitials are introduced into the system due to irradiation [27]. Fukuoka et al. [28] and Sugai et al. [29] investigated the effect of high energy Xe and Ni ion irradiation on the electrical, optical and structural properties of Al doped ZnO films. They observed an increase in conductivity of the Al doped films with ion irradiation and suggested that the irradiation induced band gap modification has close relation with the conductivity increase. Formation of single phase Co-implanted ZnO thin films using swift heavy ion (SHI) irradiation has been reported by Angadi et al. [21] and R. Kumar et al. [30]. They found a decrease in the electrical resistivity of the irradiated samples and observed close interplay between electrical and magnetic properties. Also, very recently Sunil Kumar et al. reported SHI induced modifications in Co doped ZnO thin films and concluded that SHI irradiation can be used to improve the quality of the thin films by intrinsically modifying the structural and optical properties [20]. To the best of our knowledge, there is no such report on irradiation induced modification of electrical transport in Mn doped ZnO system and so a systematic investigation has been carried out.

## 2. Experiments

The $Zn_{1-x}Mn_xO$ (x = 0.02, 0.04) samples were synthesized by conventional solid-state reaction method [6,16,31]. Stoichiometric amount of ZnO and $MnO_2$ powders (each of purity 99.99%; Sigma-Aldrich, Germany) have been weighed, mixed and ground together. The samples were initially milled for 32 h followed by annealing at 400 $^0$C for 8 h. The resulting powder was again milled for another 64 h. All the milling was performed in "Fritsch planetary mono mill" machine (Model no: Pulverisette 6) using agate ball and container. In order to avoid large grain size reduction (and hence to avoid the grain size related effects), the samples have been milled at ball to mass ratio of 1:1. The powder thus obtained was then pressed into pellets, followed by final sintering at 500 $^0$C for 12 h. The reasons for choosing 500 $^0$C as the final sintering temperature has been discussed earlier [31]. The synthesized $Zn_{0.98}Mn_{0.02}O$ and $Zn_{0.96}Mn_{0.04}O$ samples have been irradiated with 50 MeV $Li^{3+}$ ion beam. The samples have been irradiated at four different fluence of $1 \times 10^{12}$, $1 \times 10^{13}$, $5 \times 10^{13}$ and $1 \times 10^{14}$ ions/cm$^2$. The irradiation experiment was carried out using a focused beam, carefully scanned over an area of 1 cm × 1 cm, after mounting the samples on the ladder in high vacuum irradiation chamber. In order to avoid the possibility of Li implantation related effects, the irradiation experiments were performed on samples of thickness of around 200 micron, less than the penetration depth (220 micron) of 50 Mev $Li^{3+}$ ion beam in Mn doped ZnO.

The phase characterization of the $Zn_{1-x}Mn_xO$ (x = 0.02, 0.04) samples before and after irradiation has been carried out using powder x-ray diffractometer [Philips, Model: PW1830] with Cu-$K_\alpha$ radiation. All X-Ray Diffraction (XRD) measurements were carried out in the range of $20^0 \leq 2\theta \leq 80^0$ in θ-2θ geometry. The electrical resistivity as a function of temperature of all the samples was measured using conventional two-probe technique. A Keithley electrometer (model 6514) was employed to measure the resistance. Positron annihilation lifetime (PAL) measurement at RT was performed on $Zn_{0.98}Mn_{0.02}O$ sample with 0, $1 \times 10^{12}$, $5 \times 10^{13}$ ions/cm$^2$ irradiation fluence. For PAL study, a 10-μCi $^{22}$Na positron source (enclosed in 2 micron thin mylar foil) was sandwiched between two identical plane faced pellets of the samples. The PAL spectra were measured with a fast-slow coincidence assembly having 182 ± 1 ps time resolution [14]. Measured spectra were analyzed by computer program PATFIT-88 [32] to obtain the possible lifetime components $\tau_i$ and their corresponding intensities $I_i$.

### 3. Results and Discussion

Figure 1 reveals that the synthesized $Zn_{1-x}Mn_xO$ sample with x=0.02 are in single phase and no traces of impurity peak has been detected. No detectable amorphisation has been observed up to highest fleunce ($1\times10^{14}$ ions/$cm^2$) reflecting high radiation hardness of ZnO based systems. For 4 at % Mn doped un-irradiated sample (figure 2), apart from the peaks corresponding to those of ZnO hexagonal wurtzite structure, a weak (112) peak of $ZnMn_2O_4$ has been observed at $2\theta = 29.11^0$ [31]. Interestingly, this impurity peak disappeared just after irradiation with lowest fluence ($1\times10^{12}$ ions/$cm^2$), as shown in inset of figure 2. This indicates that the impurity phase has been dissolved and the sample become single phased, at least within the detection limit of XRD. Irradiation induced dissolution of impurity phase has also been observed earlier for Ag ion irradiated Co and Fe doped ZnO thin films [21,27,30]. But for Mn doped ZnO system, probably this is the first report of impurity phase dissolution using comparatively lighter ion, Li beam irradiation. The irradiated particles after entering the target suffers a number of collisions both elastic and inelastic and loses its energy via these interactions viz., electronic energy loss ($S_e$) and nuclear energy loss ($S_n$) respectively [20,21,27]. The evaluation using the simulation program SRIM-2008 [33] showed that the mean electronic energy loss $S_e$ and nuclear energy loss $S_n$ of 50 MeV $Li^{3+}$-ions in our synthesized Mn doped ZnO samples are 13.69 eV/Å and $7.65 \times 10^{-3}$ eV/Å, respectively. Thus irradiated Li ion loses its energy mainly by electronic energy loss process. Transfer of energy to the lattice, locally, can modify the impurity phase to more stable doped ZnO structure. Thus, in concurrence to the earlier reports of obtaining single phase TM doped ZnO thin films by ion beam irradiation [20,21,27,30], our study also lead to possibility of single phase formation in 4 at % Mn doped ZnO with Li ion irradiation.

Furthermore, we have monitored the variation of peak intensity as well as FWHM of the most dominant (101) peak with irradiation fluence. Figure 3 shows that both the intensity and FWHM of the (101) peak of $Zn_{0.98}Mn_{0.02}O$, decreases strongly with initial irradiation fluence and gradually saturates at higher fluence. This decrease in peak intensity along with its decreasing FWHM appears to be contradictory in nature, as far as crystallinity of the sample is concerned. But this apparent contradicting behavior has also been reported earlier in 100 KeV Ne and 1.2 Mev Ar irradiated ZnO [34,35]. It is noteworthy to mention that, we observed some peculiar behavior in temperature dependent resistivity data of irradiated $Zn_{0.98}Mn_{0.02}O$ samples, which may have some correlation with the strange features obtained in XRD for the same samples. In general, polycrystalline samples are rich in defects. Also the grain boundaries (GB) are much more defective compared to the grain interiors [36]. During the passage of energetic projectiles the re-organization of defects in the GB region is more likely due to larger abundance of defects there. Thus, with increasing fluence up to the dose of $1x10^{13}$ ion/cm$^2$, the distribution of the grain orientation becomes sharper with increase in grain size, which is reflected in the decrease in FWHM of the most intense (101) peak [35,37]. Whereas in the grain interiors, before irradiation which was less defective region, the energy lost by the high energy ion beam also creates some defects. Thus within each grain, the crystalline quality degrades with increasing ion dose, as also reported by Matsunami et al [35]. This could explain the decrease of the XRD intensity of the most dominant (101) peak with initial doses of fluence. Thus high energy ion beam irradiation affect the GB regions and grain interior regions in two mutually opposite direction, as obtained earlier by Matsunami et al [35]. With further increase of doses of irradiation, the competition between these two effects leads to homogenization of defects to some extent. Also a saturation of defects is expected in such case [34, 38]. This is reflected in figure 3, where with irradiation dose the FWHM as well as peak intensity initially deceases and then gradually saturates for the irradiated $Zn_{0.98}Mn_{0.02}O$ samples.

We have also monitored the variation of peak intensity as well as FWHM of the most dominant (101) XRD peak with irradiation fluence for $Zn_{0.96}Mn_{0.04}O$ (figure 4). For the initial irradiation fluence, unlike $Zn_{0.98}Mn_{0.02}O$, FWHM of (101) peak increases and peak intensity decreases for the $Zn_{0.96}Mn_{0.04}O$ sample. It should be remembered that, the as prepared $Zn_{0.96}Mn_{0.04}O$ sample contains impurity peak of $ZnMn_2O_4$ phase, which dissolve with the lowest dose of fluence. Thus upon irradiated with $1\times10^{12}$ ions/cm$^2$ dose of fluence the sample becomes single phase in nature. But we found that, with increasing dose of irradiation ($1\times10^{13}$ ions/cm$^2$), the FWHM of the (101) peak decreases and the peak intensity increases and no appreciable change for higher doses of irradiation ($5\times10^{13}$ ions/cm$^2$ to $1\times10^{14}$ ions/cm$^2$). This observation suggests that the crystalline quality of the 4 at % Mn doped ZnO samples becomes better with higher doses of irradiation. In the present sample the doping concentration of Mn is higher than $Zn_{0.98}Mn_{0.02}O$ and hence the resultant defective state after irradiation can be largely different. So, different trend in XRD features is not unexpected.

Close inspection of figure 1 indicates higher angle shift of (101) peak in case of $Zn_{0.98}Mn_{0.02}O$ sample just after the irradiation with fluence of $1x10^{12}$ ions/cm$^2$. But with increasing irradiation fluence, there is no further shift of the (101) peak. This might be due to release of residual strain in the system with irradiation [39].

But in case of $Zn_{0.96}Mn_{0.04}O$ (figure 2) sample XRD peaks shifts towards lower angle after irradiation. It indicates simply enhancement of lattice parameter. Unirradiated $Zn_{0.96}Mn_{0.04}O$ sample contains impurity peak of $ZnMn_2O_4$, which dissolves just with initial fluence of irradiation ($10^{12}$ ions/cm$^2$). So incorporation of Mn ions in host ZnO matrix increases with dissolution of impurity phase. Thus the observed shift of (101) peak towards lower angle (enhancement of lattice parameter) seems to quiet justified as ionic radii of $Mn^{2+}$ (0.67 Å) is higher than that of $Zn^{2+}$ (0.60 Å) [40]. There must be some higher angle shifting of (101) peak due to release of residual strain in the system with irradiation in case of $Zn_{0.96}Mn_{0.04}O$ samples also. But lower angle shifting of (101) peak due to more Mn incorporation with irradiation predominates over the earlier strain release effect.

The figure 5 represents the SEM images of $Zn_{0.98}Mn_{0.02}O$ samples, both unirradiated and irradiated with fluence of $1x10^{14}$ ions/cm$^2$ respectively. Figure 5 demonstrates that, grain size increases with irradiation in accordance with changes of FWHM values depicted in figure 3. It may due to release of strain in the system with irradiation. Figure 6 represents the SEM images of $Zn_{0.96}$ $Mn_{0.04}O$ samples, both unirradiated and irradiated with highest fluence ($1x10^{14}$ ions/cm$^2$) respectively. Here grain size shows an opposite trend i.e. decreases with ion irradiation. This observation is also corroborated with changes of FWHM values as indicated for same set of samples (figure 4). Dissolution of impurity peak with irradiation increases Mn incorporation in the host ZnO matrix. Now gradual increase of Mn ions provides retarding force on grain boundaries. If the retarding force generated is higher than the driving force for grain growth due to Zn, the movement of the grain boundary is impeded [41]. This in turn gradually decreases grain size with increasing irradiation. All the SEM micrographs show closely packed grains with no significant amount of agglomeration. Further distribution of grain size throughout samples is uniform and homogeneous.

Temperature dependent resistivity measurement has been carried out for $Zn_{0.98}Mn_{0.02}O$ and $Zn_{0.96}Mn_{0.04}O$ samples as shown in figure 7 and 8 respectively. For $Zn_{0.98}Mn_{0.02}O$, a monotonic decrease in $\rho_{RT}$ with increasing irradiation fluence has been observed [inset of figure 7(a)] with two orders of magnitude reduction due to highest fluence ($1\times10^{14}$ ions/cm$^2$). In a recent work, lowering of $\rho_{RT}$ by four orders of magnitude has been found in 1.2 MeV Ar irradiated ZnO [34]. Huge resistance loss due to irradiation by light/heavy energetic ions has also been observed by other groups [42]. The reason for change in resistivity due to irradiation, particularly in doped ZnO systems, is a matter of investigation till date [17]. The decrease of resistivity is due to increase of donors or deactivation of compensating acceptors or both. In polycrystalline samples, most of the vacancy clusters reside near the GB region as mentioned earlier. The region is devoid of carriers (depletion region) and act as a potential barrier during the transport of carriers between the grains. Increased donor density can reduce the height of the potential barrier in n type ZnO. On the other hand, recovery of a fraction of GB defects can also lower the carrier scattering at GB. At the same time, Dong et al [17] proposed that the presence of large vacancy clusters and huge oxygen vacancies are the source of reduced resistance in ion irradiated ZnO. It should be mentioned here that, electronic energy deposition can excite (and also ionize) the atoms and after de-excitation within few ps, a reorganization of local defect structure is possible. This process is more effective near the highly defective regions i.e, at the GB. We feel that resultant stable defect structure creates large oxygen vacancies ($O_V$) as dominant defects in ZnO based systems. Indeed, our PAL results (discussed later) reflect the existence of vacancy cluster in the pristine $Zn_{0.98}Mn_{0.02}O$ sample. However, XRD or PAL studies does not support further clustering of vacancies due to Li ion irradiation. Hence, it can be summarized that, the recovery of a fraction of GB defects as well as the presence of $O_V$'s due to irradiation contribute to the reduction of resistivity in these polycrystalline samples. The thermal variation of resistivity measurement of the irradiated $Zn_{0.98}Mn_{0.02}O$ sample shows an interesting behavior. Though the unirradiated $Zn_{0.98}Mn_{0.02}O$ sample is semiconducting in nature throughout the temperature range of measurement, but after irradiation it shows a metal to semiconductor transition. Most interestingly, no such transition is observed in the case of $Zn_{0.96}Mn_{0.04}O$ sample (all samples, irradiated and un-irradiated, showing semiconducting behavior in the measured temperature range). Recently, different groups reported metal-semiconductor transition in both doped and undoped ZnO system [27,21,43]. It is noteworthy to mention that, our observation is slightly different. We observed that the sample is metallic in low temperature regime and semiconducting in higher temperature. Also the sample irradiated with lowest fluence shows multiple transitions but for the samples irradiated with higher fluences ($1\times10^{13}$ ions/cm$^2$ to $1\times10^{14}$ ions/cm$^2$), only one transition has been observed. Angadi et al [21] and Nistor et al [43] have shown that, the presence of oxygen vacancies possibly give rise to metal-semiconductor transition in ZnO thin film. Further we observe that the transition temperature shifts towards lower temperature with increasing doses of irradiation [figure 7b]. Homogenization of defects with higher fluence may be responsible for the vanishing of multiple transitions and leading towards semiconducting behavior with increasing dose of irradiation.

We have also investigated the resistivity of the 4 at% Mn doped sample, $Zn_{0.96}Mn_{0.04}O$. We found that $\rho_{RT}$ of the un-irradiated $Zn_{0.96}Mn_{0.04}O$ sample is higher than that of 2 at% Mn doped sample. Interestingly, we observed a two orders of magnitude reduction in $\rho_{RT}$ value for the $Zn_{0.96}Mn_{0.04}O$ sample with initial irradiation fluence ($1\times10^{12}$ ions/cm$^2$). Such reduction of $\rho_{RT}$ with irradiation was also observed for $Zn_{0.98}Mn_{0.02}O$. But unlike $Zn_{0.98}Mn_{0.02}O$, $\rho_{RT}$ of the $Zn_{0.96}Mn_{0.04}O$ shows a small but steady increase, as the irradiation fluence is further increased from $1\times10^{13}$ ions/cm$^2$ to $1\times10^{14}$ ions/cm$^2$ (figure 8 (inset)). Since the $Zn_{0.96}Mn_{0.04}O$ sample contains higher percentage of Mn, the equilibrium defective state (after irradiation) of the sample is different than that of $Zn_{0.98}Mn_{0.02}O$. As $\rho_{RT}$ value is very high, we have not attempted to measure the low temperature resistivity of the un-irradiated $Zn_{0.96}Mn_{0.04}O$ sample. However, we have measured the thermal variation of resistivity of the all the irradiated $Zn_{0.96}Mn_{0.04}O$ samples (figure 8). It is noteworthy to mention that, we have been able to measure the resistivity for these of samples only down to 170 K, below which resistivity tends beyond the limit of the instrument. All the irradiated $Zn_{0.96}Mn_{0.04}O$ samples remain semiconducting down to the lowest temperature measured.

PAL measurements on un-irradiated and few irradiated $Zn_{0.98}Mn_{0.02}O$ samples show interesting features. The possible impurity phase related problem may complicate the PAL data for 4 at% Mn doped samples and so the measurement for this sample is avoided. The results of PAL spectrum analysis are shown in table 1. All the PAL spectra are found to be best fitted with three lifetime components. The longest lifetime component ($\tau_3$, 1400 ps with intensity 3-5%) originates due to the annihilation of positron from orthopositronium atoms. Decay of orthopositronium into parapositronium through pickoff annihilation gives rise to such a large lifetime. In polycrystalline samples, there always exist micro voids where orthopositronium formation is favourable [14]. As the first lifetime component ($\tau_1$) and the intermediate one ($\tau_2$) have changed significantly with irradiation fluence, we feel that both have defect related origin. $\tau_2$ and $\tau_1$ provide a qualitative indication about the spatial extension of the defects (i.e, defect size). The corresponding intensities ($I_1$ and $I_2$ respectively) reflect the relative abundance of such defect sites. For unirradiated sample the value of $\tau_1$ is close to the positron lifetime at zinc vacancies in ZnO. Indeed, Wang et al [44] have attributed the origin of $\tau_1$ from the diffused zinc vacancies at the interface of the grains. The value of $\tau_2$ indicates that there exist large vacancy clusters in the sample, possibly of type $V_{Zn-O}$ divacancies near the GB region [14]. High concentration of open volume defects and related disorder can produce foam like structure near the GB as predicted by Straumel et al. [8]. Such grain boundary disorder is drastically modified due to $Li^{3+}$ irradiation. Compared to unirradiated sample, $\tau_1$ sharply decreases after initial fluence of $1 \times 10^{12}$ ions/cm$^2$. $\rho_{RT}$ also shows decrease with increasing irradiation fluence. This is due to formation of increasing neutral oxygen vacancies with irradiation fluence. As decreasing trends of $\tau_1$ as well as $\rho_{RT}$ with irradiation indicate increasing presence of free electrons in the system. It is possible that a fraction of the injected positrons are now annihilated in the bulk of the sample. So $\tau_1$ becomes closer (but lower) [45] to the bulk lifetime of positrons in ZnO after irradiation. It is indeed an interesting observation. $\tau_2$ also shows significant reduction due to initial dose of irradiation. Hence, we feel that there is no evidence of increase in vacancy clusters with initial dose of irradiation. Also, we observe that, $\tau_1$ and $\tau_2$ shows a minute increase with further irradiation fluence. However, $\tau_1$ and $\tau_2$ are fitting dependent parameters to some extent. In such complex systems, statistically more accurate parameter is average positron lifetime, $\tau_{av} = (\tau_1 I_1 + \tau_2 I_2)/(I_1+I_2)$, which reflects the overall defective nature of the sample [14,46]. $\tau_{av}$ decreases due to irradiation with $1 \times 10^{12}$ ions/cm$^2$ flunce and remains unaffected due to further increase of irradiation.

## 4. Conclusion

The un-irradiated and 50 MeV $Li^{3+}$ ion irradiated $Zn_{1-x}Mn_xO$ (x = 0.02, 0.04) samples were characterized by XRD, temperature dependent resistivity and room temperature PAL spectroscopy.

XRD result indicates wurtzite type structure for $Zn_{0.98}Mn_{0.02}O$ but impurity (112) peak of $ZnMn_2O_4$ was developed apart from usual peaks of ZnO for $Zn_{0.96}Mn_{0.04}O$. Ion irradiation dissolves the impurity phase. SEM micrographs indicate homogeneity of the samples with uniform particle size. Room temperature resistivity values decrease abruptly with irradiation. Generation of oxygen vacancy, Zn interstitial and recovery of a part of defects at GB due to irradiation may be the reason of resistivity reduction. The temperature dependent electrical resistivity results for irradiated $Zn_{0.98}Mn_{0.02}O$ samples shows shifting of metal-semiconductor transition temperature towards lower side with increasing fluence of irradiation. The $\rho_{RT}$ values of the $Zn_{0.96}Mn_{0.04}O$ are higher compared to that of $Zn_{0.98}Mn_{0.02}O$ samples. No metal-semiconductor transition is observed in any irradiated $Zn_{0.96}Mn_{0.04}O$ samples from the temperature dependent of resistivity measurement at least down to 170

K. PAL spectroscopy data analysis indicates that increasing fluence of irradiation cause lowering of defects in $Zn_{0.98}Mn_{0.02}O$. This fact is also supported from XRD and electrical resistivity measurements.

**Acknowledgement**

The authors are thankful to IUAC, New Delhi for providing the ion beam irradiation facilities. One of the authors (SB) is also thankful to Department of Science and Technology (DST), Govt. of India and IUAC, New Delhi for providing financial support in the form of sanctioning research project, vide project no.: SF/FTP/PS-31/2006 and UFUP-43308 respectively. Author SKN is thankful to University Grants Commission (UGC) for providing him Junior Research Fellowship and author SC is grateful to Government of West Bengal for providing financial assistance in form of University Research Fellowship.

**Figure Captions:**

**Figure 1:** XRD of un-irradiated and irradiated 2 at% Mn doped ZnO sample.

**Figure 2:** XRD of un-irradiated and irradiated 4 at% Mn doped ZnO sample. Inset: Enlarge view of XRD in the range $28^o$ to $30^o$.

**Figure 3:** Variation of FWHM and intensity of (101) peak with irradiation fluence for 2 at% Mn doped ZnO sample.

**Figure 4:** Variation of FWHM and intensity of (101) peak with irradiation fluence for 4 at% Mn doped ZnO sample.

**Figure 5:** SEM micrographs of $Zn_{0.98}$ $Mn_{0.02}O$ samples, (a) unirradiated and (b) irradiated with fluence of $1x10^{14}$ $Li^{3+}$ ions/$cm^2$.

**Figure 6:** SEM micrographs of $Zn_{0.96}$ $Mn_{0.04}O$ samples, (a) unirradiated and (b) irradiated with fluence of $1x10^{14}$ $Li^{3+}$ ions/$cm^2$.

**Figure 7:** (a) Thermal variation resistivity of un-irradiated $Zn_{0.98}Mn_{0.02}O$ sample. Inset: Variation of $\rho_{RT}$ of $Zn_{0.98}Mn_{0.02}O$ with irradiation irradiation fluence. (b) Thermal variation resistivity of all irradiated $Zn_{0.98}Mn_{0.02}O$ sample.

**Figure 8:** Thermal variation resistivity of all irradiated $Zn_{0.96}Mn_{0.04}O$ sample. Inset: Variation of $\rho_{RT}$ of $Zn_{0.96}Mn_{0.04}O$ with irradiation fluence.

Table 1: The fitting parameters found from positron annihilation lifetime measurement on $Zn_{0.98}Mn_{.02}O$ samples irradiated with different irradiation fluence

| Irradiation fluence (ions/cm$^2$) | $\tau_1$ (ps) | $I_1$ (%) | $\tau_2$ (ps) | $I_2$ (%) | $\tau_3$ (ps) | $I_3$ (%) | $\tau_{av}$ (ps) |
|---|---|---|---|---|---|---|---|
| 0 | 222 ± 1 | 59.3 ± 0.01 | 401 ± 2 | 37.5 ± 0.01 | 1406 ± 40 | 3.2 ± 0.01 | 291 ± 3 |
| 1 ×10$^{12}$ | 167 ± 1 | 37.3 ± 0.01 | 335 ± 2 | 57.8 ± 0.01 | 1530 ± 44 | 4.9 ± 0.01 | 269 ± 3 |
| 5 × 10$^{13}$ | 188 ± 1 | 48.1 ± 0.01 | 350 ± 2 | 48.4 ± 0.01 | 1772 ± 52 | 3.5 ± 0.01 | 269 ± 3 |

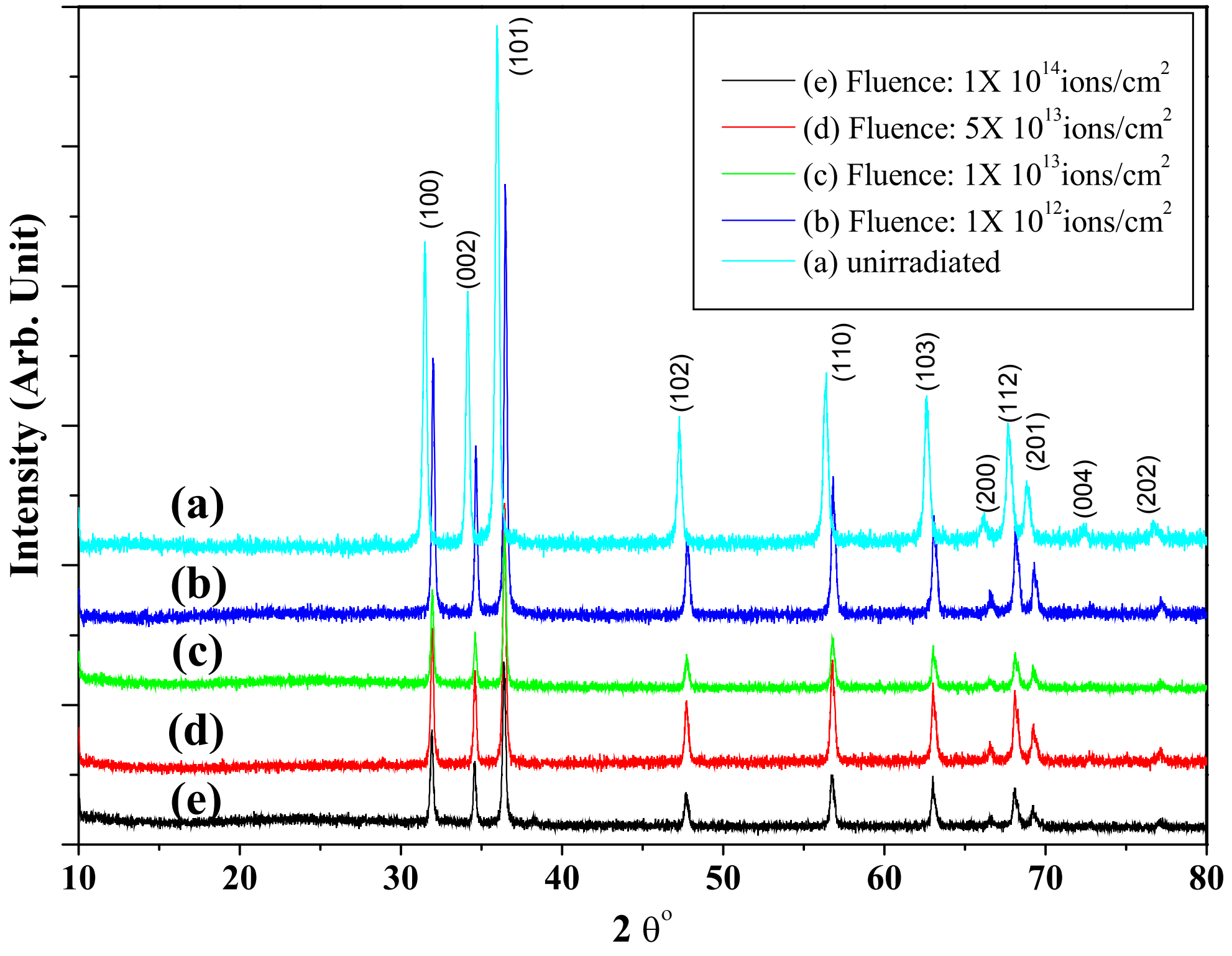

Figure 1

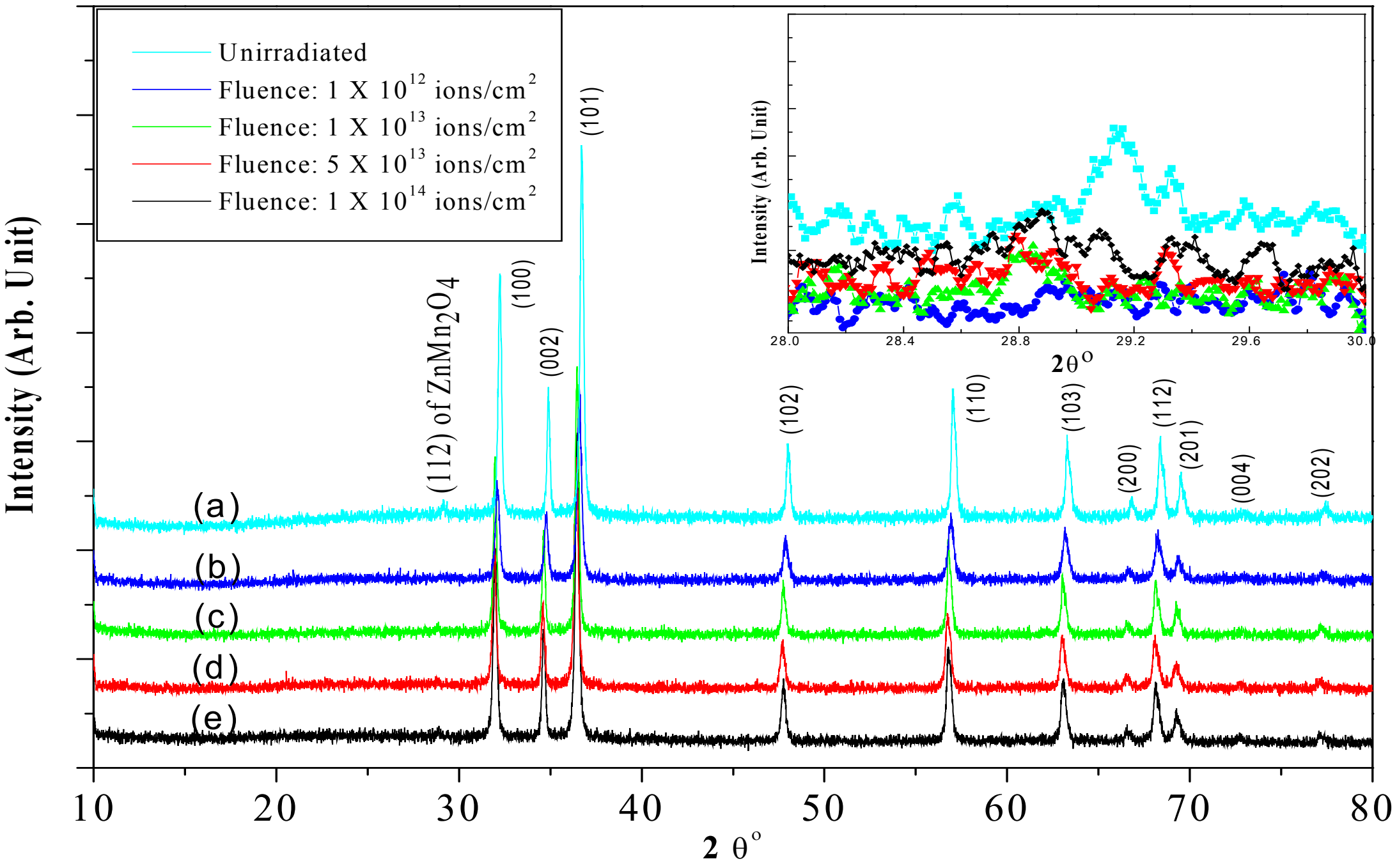

Figure 2

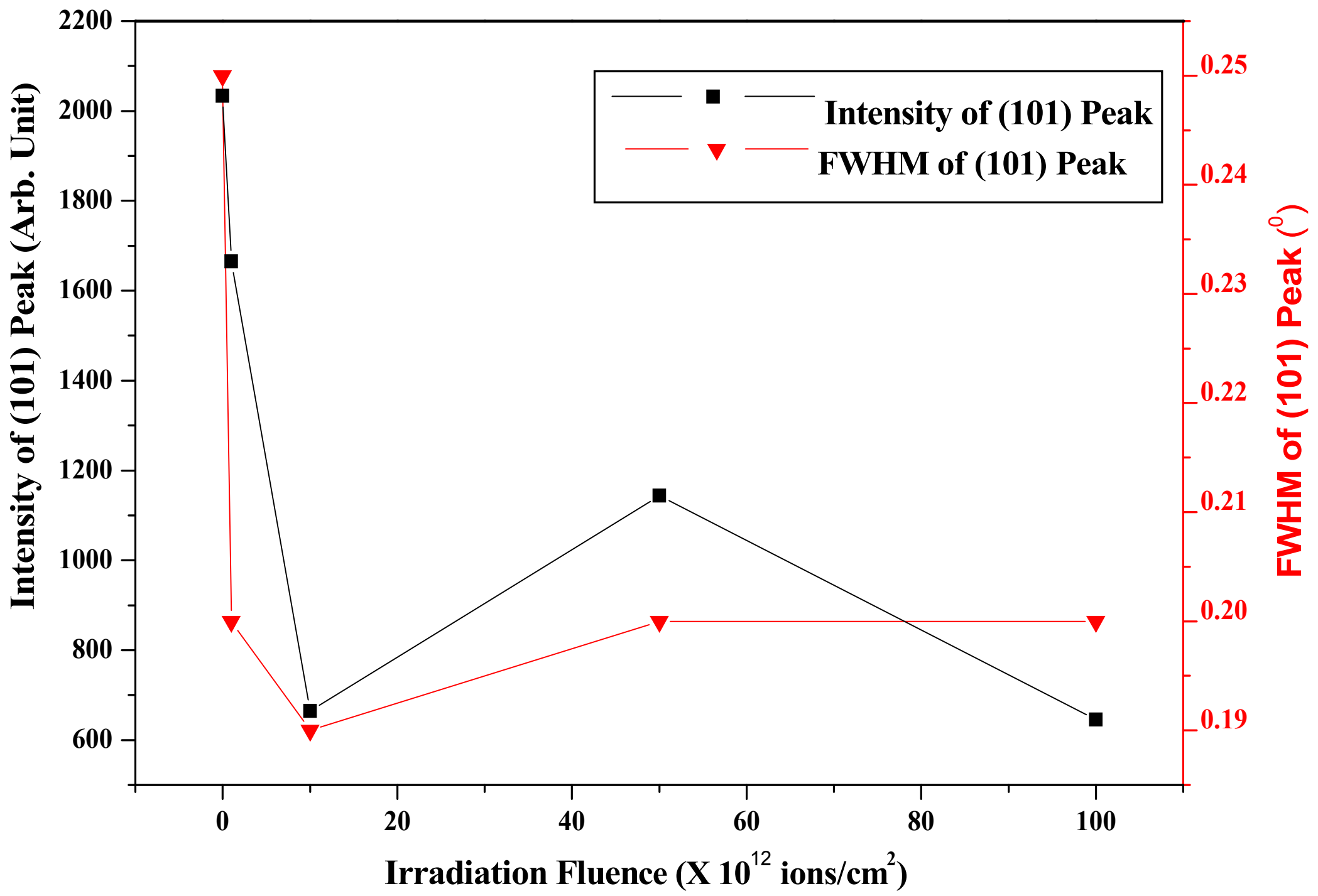

Figure 3

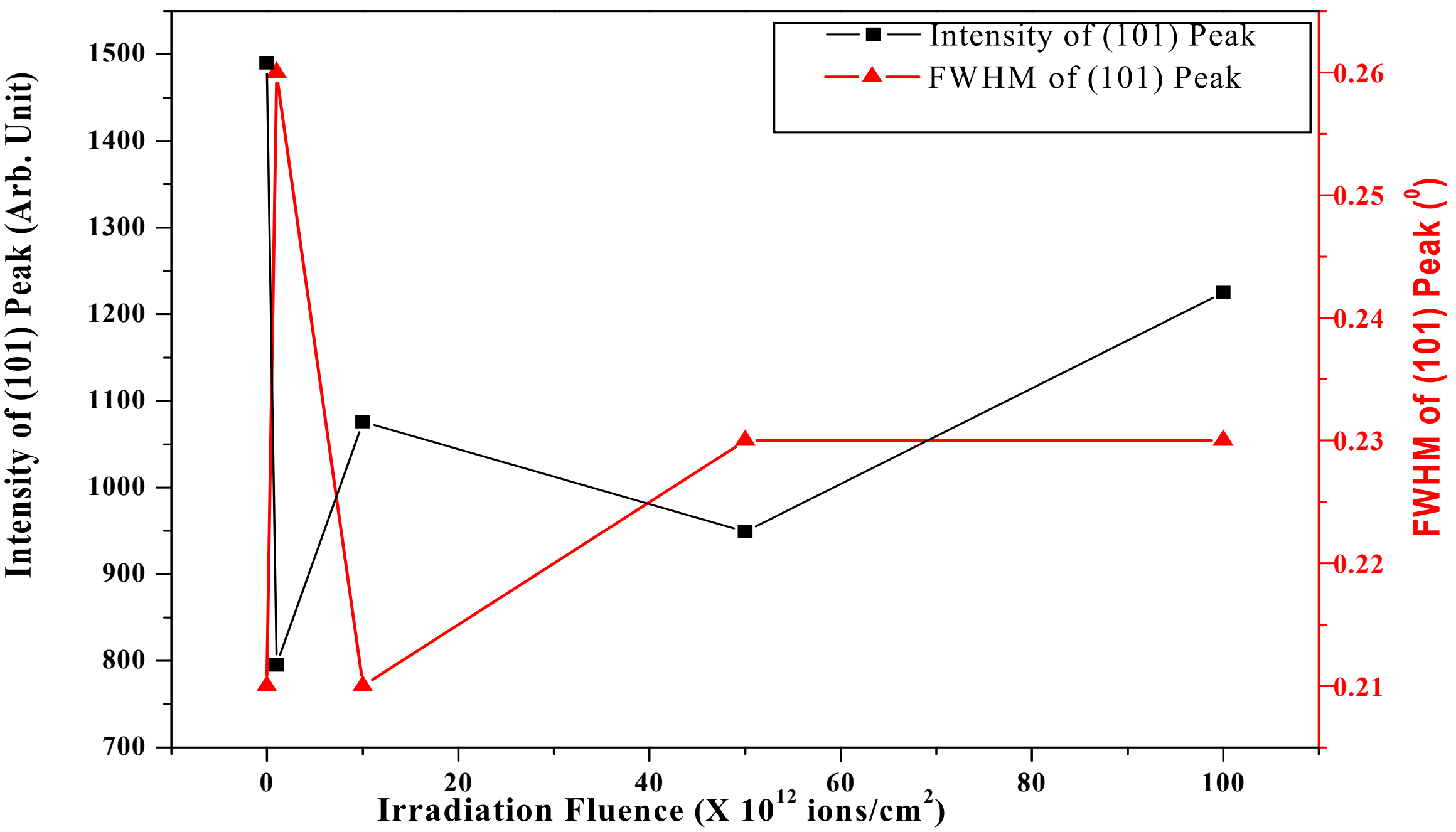

Figure 4

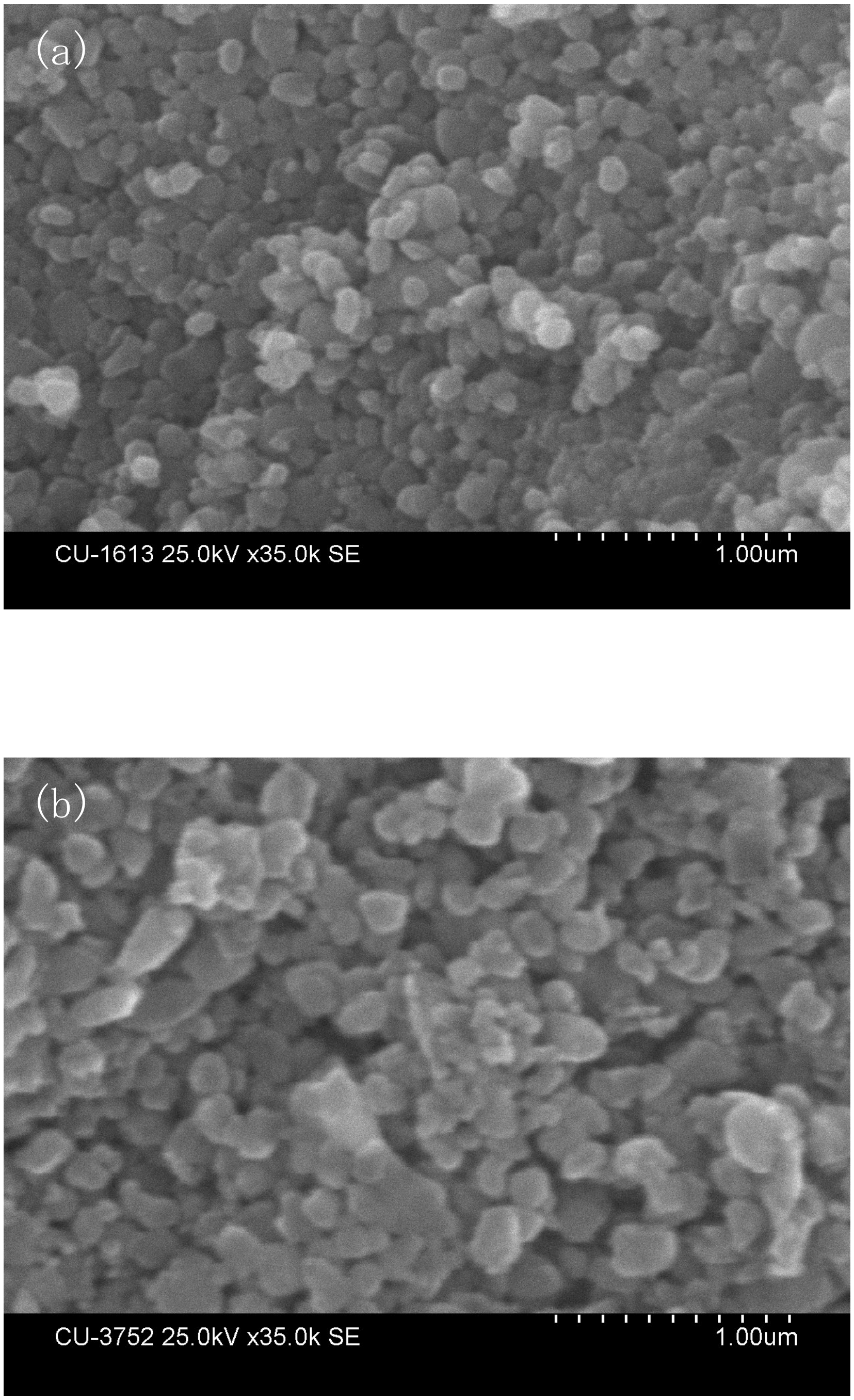

Figure 5

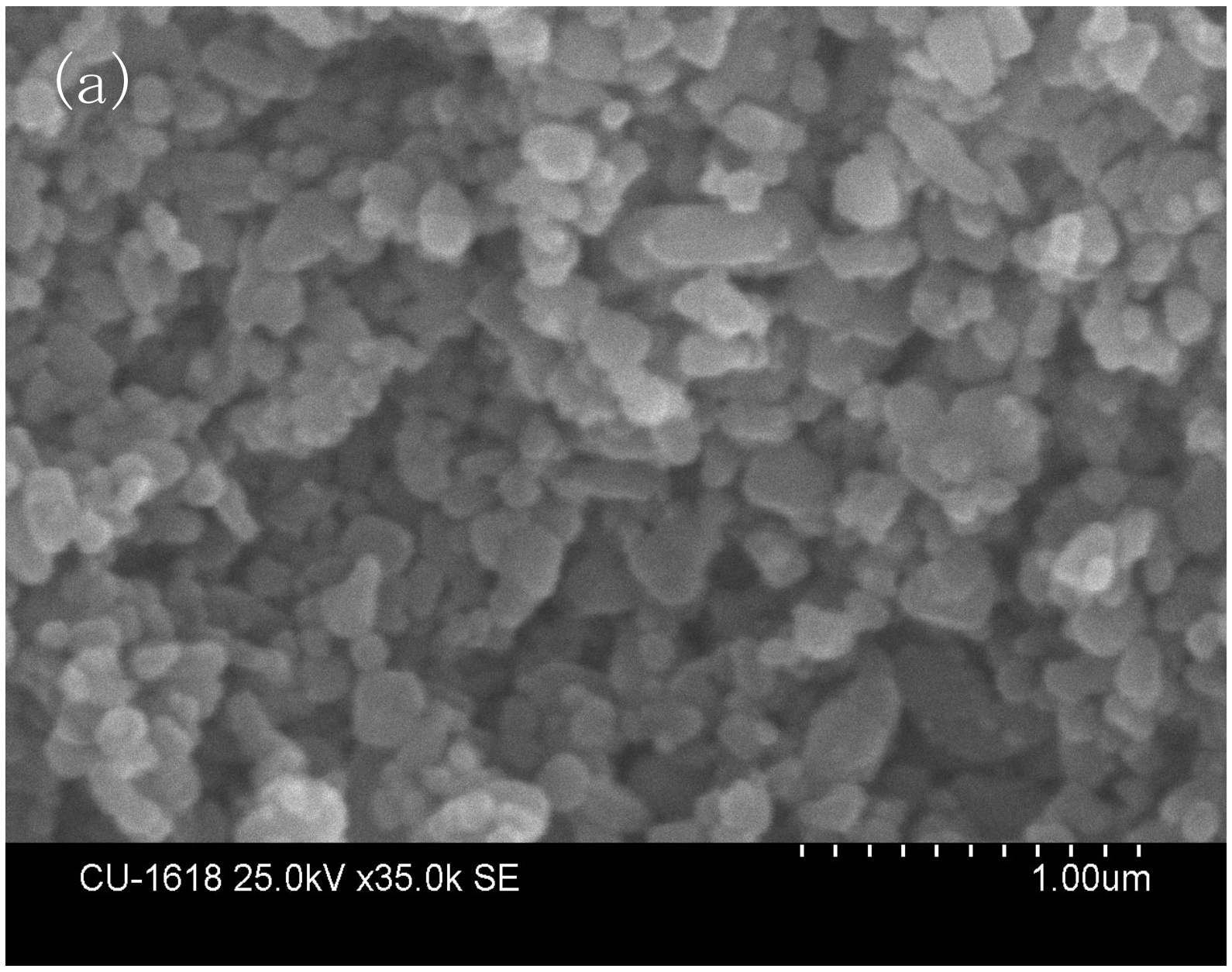

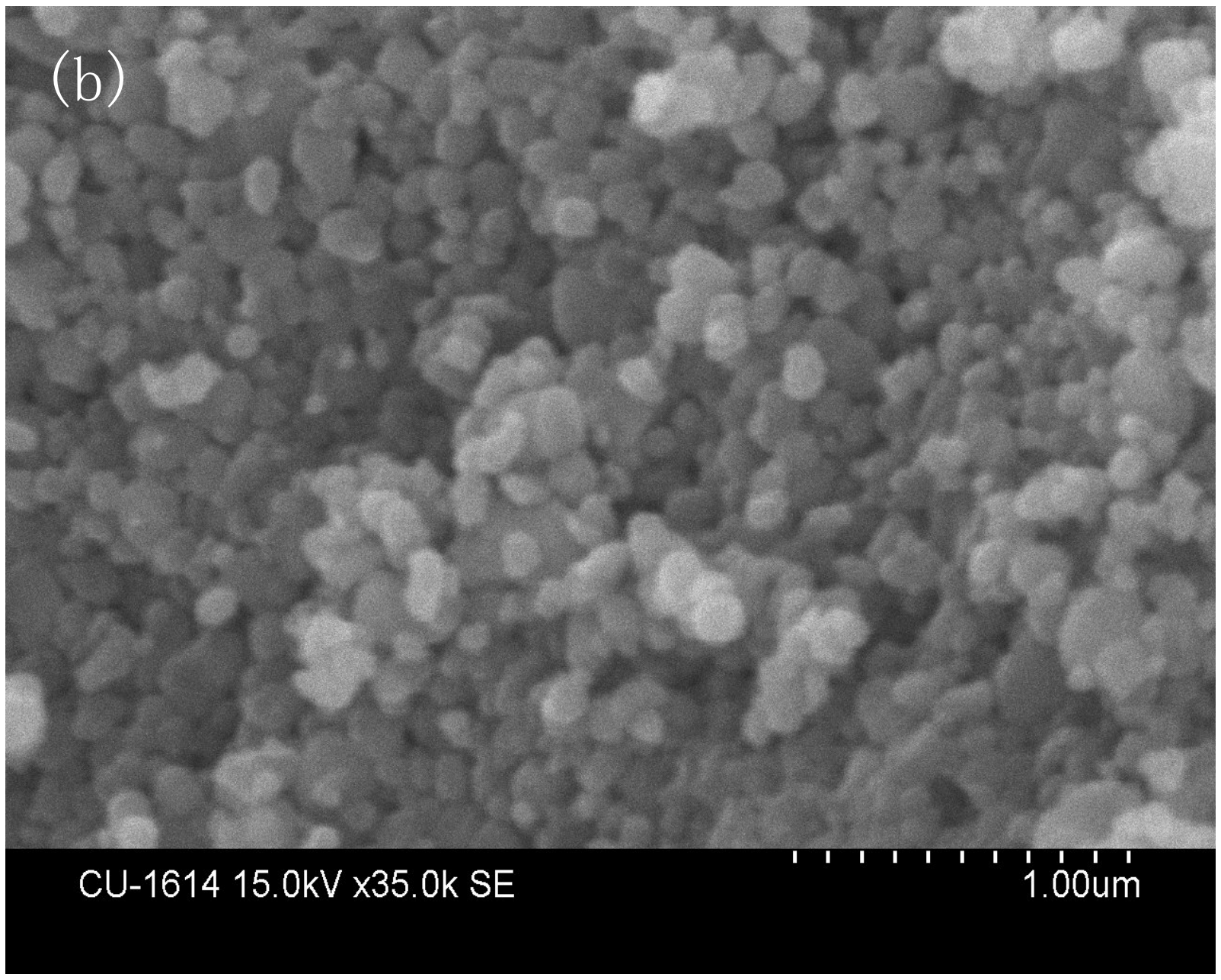

Figure 6

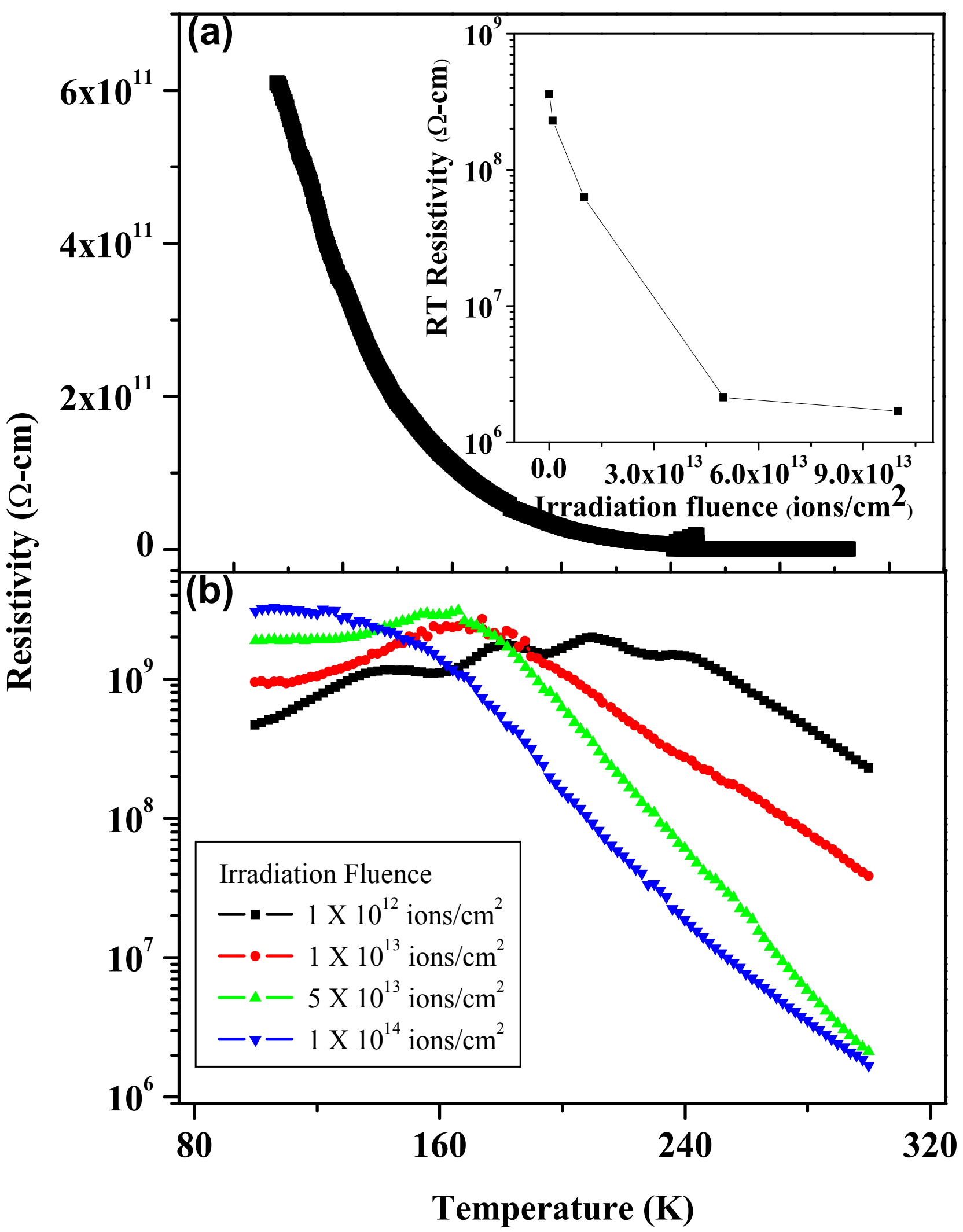

Figure 7

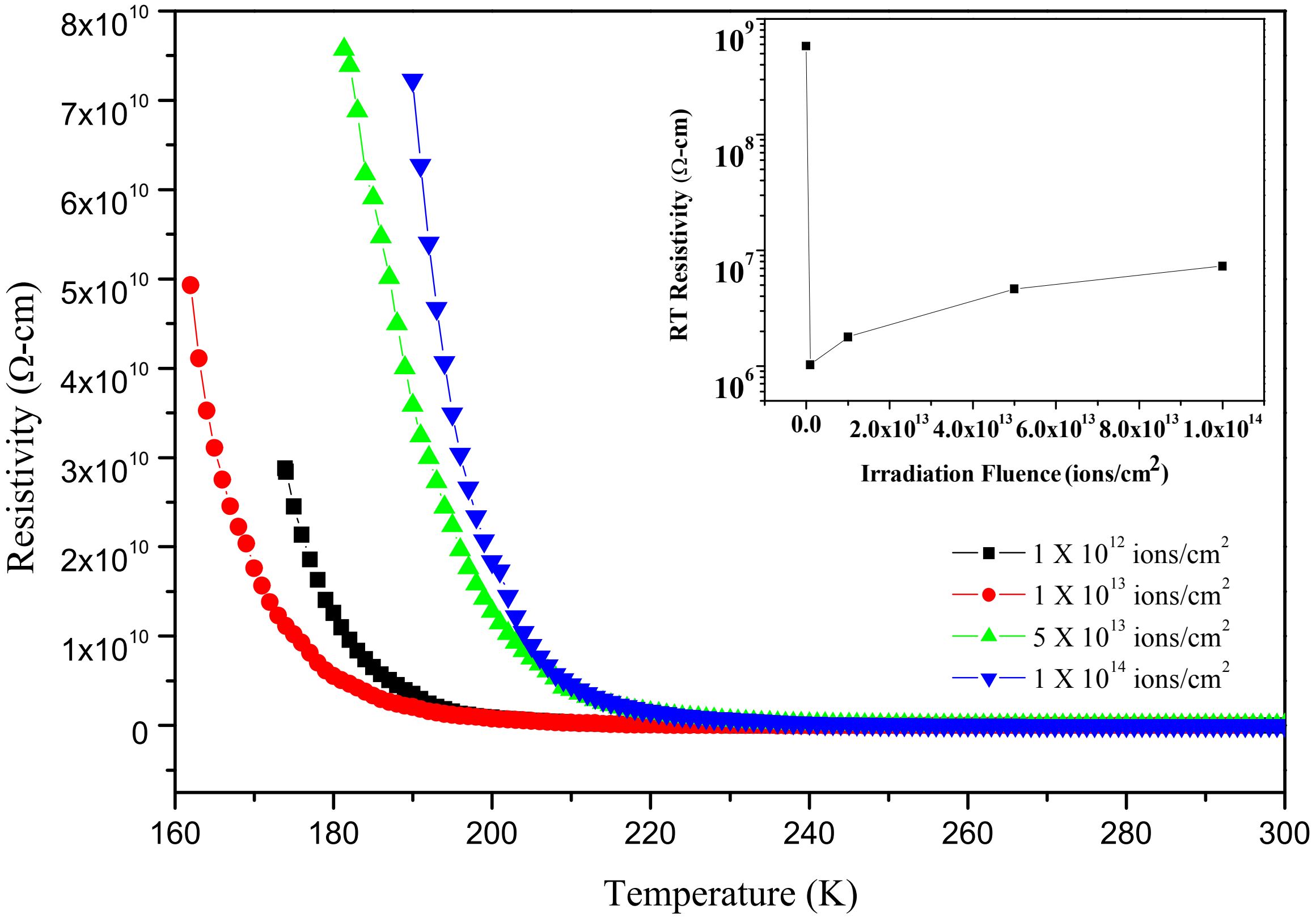

Figure 8